# Ultrafast laser pulses to detect and generate fast thermo-mechanical transients in matter

C. Giannetti[1], F. Banfi[1], D. Nardi[1,2], G. Ferrini[1], F. Parmigiani[3,4]

[1]Department of Mathematics and Physics, Università Cattolica, I-25121 Brescia, Italy
[2]Università degli Studi di Milano, I-20100 Milano, Italy
[3]Department of Physics, Università degli Studi di Trieste, I-34127 Trieste, Italy
[4]Sincrotrone Trieste, I-34012 Basovizza, Trieste, Italy
Corresponding author: c.giannetti@dmf.unicatt.it

**Abstract:** The use of femtosecond laser pulses to impulsively excite thermal and mechanical transients in matter has led, in the last years, to the development of picosecond acoustics. Recently, the pump-probe approach has been applied to nano-engineered materials to optically generate and detect acoustic waves in the GHz-THz frequency range. In this paper, we review the latest advances on ultrafast generation and detection of thermal gradients and pseudo-surface acoustic waves in two-dimensional lattices of metallic nanostructures. Comparing the experimental findings to the numeric analysis of the full thermo-mechanical problem, these materials emerge as model systems to investigate both the mechanical and thermal energy transfer at the nanoscale. The sensitivity of the technique to the nanostructures mass and shape variations, coupled to the phononic crystal properties of the lattices opens the way to a variety of applications ranging from hypersonic waveguiding to mass sensors with femtosecond time-resolution.

## 1. Introduction

The use of ultrafast laser pulses (pulse duration <1 ps) to generate and detect thermo-mechanical transients in matter opened the field of picoseconds ultrasonics [Maris1998]. The basic idea is to use sub-ps light pulses in a "percussion" approach: an intense pulse (pump), focused on an area A of a solid surface, impulsively delivers an energy Q within the light penetration depth $\xi$, inducing a non-equilibrium local heating of both electrons and lattice on the ps-timescale. The local temperature increase $\Delta T = [Q/(A\xi C)] \cdot e^{-z/\xi}$ (C being the specific heat per unit volume) is coupled to a sudden lattice expansion through the thermal expansion coefficient α. The photoinduced thermoelastic stress is able to launch strain pulses $\eta(z,t)$ propagating away from the excited surface along the direction z, regulated by [Maris1986]:

$$\eta(z,t) = \frac{Q\alpha}{A\xi C}\frac{1+\nu}{1-\nu}f(z-vt) \tag{1}$$

ν being the Poisson ratio and *f(z-vt)* the function describing the shape of the strain pulse, moving at the longitudinal sound velocity v. The dependence of the refractive index on the strain, through the photoelastic constant, allows following the propagation of strain pulses by means of a second delayed pulse (probe). An energy per pulse of the order of Q~1 nJ, easily available by means of Ti:sapphire oscillators producing ~100 fs light pulses, can be exploited to impulsively heat semiconductor or metal samples leading to temperature raises ΔT~0.1-10 K, implying thermoelastic stresses ranging from 0.1 to 1 Mbar.

The first evidences of picosecond acoustic transients in semiconductors and metals, reported in the seminal work by H. Maris et al. [Maris1986], triggered the investigation of the microscopic mechanisms responsible for the strain generation. At present, there is general agreement that, in semiconductors, the accumulation of the long-lived photo-injected excitations modifies the orbital population, inducing an electronic stress coupled to an impulsive contraction or expansion of the lattice through the deformation potential [Morozov2008]. This mechanism is always accompanied by the volume expansion related to the increase of the phonon population through the anharmonicity of the crystal potential. The latter effect, named lattice thermoelasticity, completely dominates in metals, where a rapid intraband relaxation of high-energy electrons is achieved by means of electron-phonon scattering. The temperature dynamics of the photoinduced non-equilibrium electrons and phonons in metals can be satisfactorily described by the two-temperature model (2TM) [Kaganov1957]. The main prediction is a local quasi-thermalization at an effective temperature T* of electrons and phonons, on the timescale of ~1 ps. The short thermalization time ensures that part of the energy Q of the pump pulse is efficiently converted into strain energy. This characteristic made very common the use of metal thin films as light-sound transducer to launch ultrasonic acoustic pulses in materials [Rogers2000, Pezeril2008].

The use of high-energy amplified Ti:sapphire oscillators opened the way to the investigation of the high-intensity strain regime, where the non-linear effects become effective. Pump pulses with energy of the order of Q~10-100 nJ generate acoustic pulses with strain amplitude of the order of $10^{-3}$ and a time duration of the order of 10 ps. In this regime the balance between the increase of the sound velocity with strain and the decrease of the velocity of the high-frequency components of the strain pulse, due to the flattening of the phonon dispersion, results in the generation of sub-THz acoustic solitons, described by a Korteweg–de Vries equation [Hao2001]. After the first evidences of acoustic solitons generation in semiconductor and transparent media [Hao2001], solitons have been exploited to modulate the band gap of semiconductor quantum-wells, through the deformation potential [Scherbakov2007].

When exciting samples at surfaces, a particular class of solutions of the elastic equation is constituted by Surface Acoustic Waves (SAWs), i.e. acoustic waves confined at the surface within a depth of the order of the wavelength λ [Landau1986]. SAWs in the hypersonic frequency range (>1 GHz) are currently used to manipulate electrons in semiconductor devices [Cecchini2005, Cecchini2006] and photons in microcavities [deLima2005, deLima2006, Rudolph2007]. The quest for efficient SAWs generation and detection techniques at even higher frequencies led to the investigation of SAWs generation by means of ultrashort laser pulses. In particular, when the pump pulse is focused on a small area (A~1-10 μm$^2$) of a surface, the large Fourier spectrum allows launching SAWs at different k-wavevectors. Time-resolved imaging techniques have been employed to follow on the ps timescale SAW propagation on free surfaces [Sugawara PRL2002, Tachizaki2006], through grain-boundaries [Hurley2006], in phononic crystals [Profunser2006] and in resonators [Maznev2009].

The possibility to selectively excite bulk quasi-monochromatic acoustic waves became real by employing engineered materials with artificial periodicities of the elastic properties. In these systems, the pump pulse induces an impulsive strain with the wavelength matching the periodicity of the structure. The use of semiconductor superlattices allowed exciting coherent THz acoustic phonons [Bartels1999] via impulsive stimulated Raman scattering and, only recently, to access the dynamics of hypersonic acoustic pulses confined into an acoustic nanocavity [Huynh2006].

Parallel efforts have been devoted to the use of ultrafast light pulses to excite acoustic eigenmodes of confined nanostructures, after the first study of the vibrational modes of gold nanostripes on a fused quartz substrate [Lin1993]. Recently, advances in the preparation of quasi-monodispersed nanostructures enabled, employing picosecond ultrasonic techniques, to investigate photo-induced coherent oscillations in metallic nanoparticles [Nisoli1997, DelFatti1999], nanocolums [Burgin2008b] and nanoprisms [Huang2004, Huang2005, Taubert2007, Burgin2008a], achieving optical control of the acoustic vibrations [Arbouet2006]. More recently, the feasibility of pump-probe measurements on single metallic nanoparticles [Burgin2008a], exploiting the surface plasmon resonance, opened the way to investigate the mechanical properties of single nanostructures, whose shape and interaction with the environment can be exactly determined by means of electron microscopy techniques [Billaud2008].

Notably, the same pump probe technique can be applied to study heat transport in matter [Stoner1993]. The pump-induced ΔT triggers a heat flow on the sub-ns timescale. The dependence of the refractive index on the temperature enables following the propagation of heat pulses by means of the probe pulse. This technique, named time-domain thermoreflectance, has been employed to investigate the thermal conductance at metal-metal [Gundrum2005] and metal-dielectric [Stoner1993, Lyeo2006] interfaces and to disentangle the energy transport related to electron diffusion [Gundrum2005], anharmonic phonon decay [Lyeo2006] and ballistic phonon transport [Highland2007]. Recently, the signature of ballistic heat transport [vonGutfeld1964], at cryogenic temperatures, has been reported in a GaAs crystal covered by a metallic thin film transducer [Perrin2006]. The extension of this technique to the study of thermal transport between a single metallic nanoparticle and the environment is a more difficult task, due to the difficulties in controlling the properties of the nanoparticle-environment interface [Voisin2000].

The frontier in this intriguing research field is the investigation of the thermo-mechanical transients in lattices of metallic nanostructures on surfaces [Lin1993, Hurley2002, Giannetti2007, Robillard2007, Hurley2008]. State-of-the-art nano-lithography and patterning techniques allow obtaining metallic nanostructures, whose shapes, dimensions, periodicities and interface properties can be carefully tuned. The interest in these systems is inherent to the following features: i) the periodicity, scalable down to ~50 nm [Chao2005], that can be exploited to launch quasi-monochromatic SAWs in the substrate at ~100 GHz [Siemens2009]; ii) the periodicity of the elastic properties, that induces the opening of a band gap in the acoustic modes [Nardi2009] (hypersonic phononic crystals); iii) the fine control over the nanostructures/substrate interface, mandatory to study heat transport at the nanoscale.

## 2. Ultrafast generation and detection of SAW in nanostructured surfaces

The scheme of the experimental set-up used to generate and detect mechanical transients in nanostructured arrays, in the diffraction configuration [Giannetti2007], is reported in Figure 1. The laser source is a Ti:sapphire oscillator, delivering short pulses with 120 fs time duration, 30 nJ energy/pulse and 790 nm wavelength, at a repetition rate of 76 MHz. The output is split into an intense component (pump) and a weak component (~1 nJ, probe). A delay line in a double-pass configuration is used to change the delay *t* between the pump and probe pulses. Two spatial points of the pump beam are imaged onto two quadrant photodiodes interfaced to a feedback system that drives two piezo-nanomotors mounted on optical mirrors. This system is used to keep the spatial position of the pump beam fixed during the experiment and avoid variations of the detected signal due to the pump-probe misalignment. In this configuration, a difference in the optical path of ~1 m can be achieved, corresponding to a temporal time-window of 3 ns. To optimize the optical signal-to-noise ratio, the pump and probe beams are focused on the same point of the sample (see Figure 2) with a size of 60 μm and 40 μm, respectively. This size allows exciting and probing a large number of nanostructures (~3·$10^3$), while keeping the laser fluence high enough to significantly excite the system. A Peltier device is used to keep the sample back side temperature $T_0$ constant during the experiment. The pump beam intensity is modulated at 100 kHz by a Photo-Elastic Modulator (PEM) placed between two crossed polarizers. The first-order diffracted beam is detected by a photodiode and filtered by a lock-in amplifier referenced at the PEM frequency. In this configuration, the measurement of the probe intensity variation $\Delta I_{1D}/I_{1D}(t)$, induced by the interaction on the sample with the pump pulse, is performed on the diffracted beam to increase the signal-to-noise ratio [Giannetti2007] and avoid all non-periodic contributions, such as Brillouin scattering from the acoustic pulses propagating into the substrate. Relative intensity variations of the order of $10^{-6}$ can be measured with a time resolution of ~120 fs on a time-window of 3 ns.

The high-frequency modulation of pump intensity has two main outcomes. The first is that the 1/f noise is strongly reduced in the lock-in detection, the second is related to the variation of the local effective temperature $T_{eff}$ of the sample excited by the modulated pump beam. Considering the rate equation governing the temperature relaxation of the system excited at t=0 by the absorbed power per unit volume W(t)= [Q/(Aξ)]·(rep.rate) [Giannetti2007], we obtain two different regimes (see Figure 3). If the modulation period $t_{mod}$ is larger than the time $t_{ss}$ needed for the system to reach the steady state temperature, the local effective temperature is $T_{eff}(t) \approx T_0 + (W(t)/C) \cdot t_{ss}$, following the pump modulation. On the contrary, if $t_{mod} < t_{ss}$ the local effective temperature is given by [Giannetti2007]:

$$T_{eff}(t) = T_0 + T_{av} + \frac{1}{C}\int_0^t W(t')dt' \tag{2}$$

In this case, a damped and π/2-dephased modulation of the effective temperature is superimposed to an average effective temperature $T_{av}$. Considering the specific heat and the thermal conductance of the silicon substrates used in the experiments, we obtain $t_{ss}$~1 ms, significantly larger than the $t_{mod}$=10 µs. For this reason, acquiring the 100 kHz in-phase and the π/2 out-of-phase components of the probe signal, we are able to decouple the genuine variation of the diffracted intensity on the sub-ns timescale, from the average heating of the substrate. On the contrary, standard modulation of the pump beam at 0.1–1 kHz would result in a high in-phase background in the probe signal, drastically decreasing the signal to noise ratio.

Two-dimensional square lattices of Permalloy (Py, $Fe_{20}Ni_{80}$) nanodisks, have been prepared through electron-beam lithography and lift-off techniques on a Si(100) surface (see Figure 2). The time-resolved measurements of the diffracted intensity have been performed on two samples with periodicity P=(1000±10) nm, diameters $d_1$=(630±10) nm and $d_2$=(400±10) nm and thicknesses $h_1$=(30±2) nm and $h_2$=(50±2) nm, respectively. The array periods, the nanodisk thicknesses and diameters have been carefully measured by atomic force microscopy (AFM).

In Figure 4a, we report the time-resolved measurements performed on the 30 nm-thick sample, with a pump energy of the order of 10 nJ/pulse. At t=0 a fast increase of the transient signal is measured, while a nanosecond decay superimposed to a faster oscillation is detected for positive delays. Considering the laser energy density absorbed by the nanostructures and by the substrate and the specific heat of both Py and Si, we can estimate that, within 5 ps, the temperature of the nanodisks is homogeneously increased by ~10 K, whereas the substrate temperature is essentially unvaried due to the different penetration depth of the 800 nm radiation [Giannetti2007]. The impulsive (t=5 ps) temperature mismatch ΔT triggers a non-equilibrium expansion of the nanostructures diameters ($\delta d/d = \alpha^* \Delta T \cdot 2 \cdot 10^{-5}$, $\alpha^*$ being the effective thermal expansion coefficient of the Py/Si system), with a periodicity given by the periodicity of the lattice. At positive delays (t>5 ps) the periodic strain induced in the substrate significantly overlaps to a pseudo-SAW acoustic eigenmode [Nardi2009]. We stress the fact that the measured signal is dominated by the pseudo-SAW contribution as far as the thickness of the nanostructures is of the order of ξ (ξ~20-50 nm in metals) and a homogeneous excitation is induced. In the regime h>>ξ, we would expect a dominant contribution from the propagation of the acoustic strain pulses within the nanostructures.

In the h≤ξ scenario, the size of the nanostructures oscillates around an equilibrium value proportional to the average temperature. The average diameter d decays as long as the heat exchange with the substrate is effective. The time dependence of the dot size is conveniently mimicked by a harmonic oscillator equation with a varying equilibrium position [Giannetti2007]:

$$\frac{\delta d(t)}{d} \propto \frac{\alpha \delta T_0 \omega_0^2}{\omega_0^2 + 1/\tau^2 - 2\gamma/\tau} \left( e^{-t/\tau} - e^{-\gamma t}\cos\omega t + \frac{\beta}{\omega}e^{-\gamma t}\sin\omega t \right) \qquad (3)$$

$\delta T_0 = \delta T$(t=5 ps) being the temperature difference between the disks and the substrate at the end of the electron-phonon thermalization process, $\tau$ the time constant regulating the heat exchange between the nanodisks and the substrate, $\omega_0$ the oscillation frequency of the undumped oscillator, $\omega=(\omega_0^2-\gamma^2)^{1/2}$ the renormalized frequency, $\gamma$ the damping constant of the mode and $\beta=1/\tau-\gamma$. In Figure 4a we report the fit to the data (black line) of the sum of function (3) (red line) and a simple exponential decay ($e^{-t/\tau}$, blue dashed line). The latter term accounts for the dependence of the refractive index on the average temperature of the metal nanostructures.

In Figure 4b we report similar measurements on a lattice of nanodisks of thickness 50 nm and diameter of 400 nm. In this case, we observe an increase of both the heat exchange time $\tau$ and the oscillation period $2\pi/\omega$. In the present experiment, the heat-exchange time constant is proportional to the thickness of the nanostructures: $\tau \approx hC_{Py}\rho_{th}$ ($\rho_{th}$ is the interface thermal resistivity), as will be addressed in the next section. The predicted ratio $\tau_2/\tau_1=h_2/h_1$ perfectly matches our results. The slight increase of the oscillation period is due to the mass loading effect, i.e. as the loading of a free surface is increased, the frequency of the SAW decreases [Landau1986, Auld1990].

Our results show that the time-resolved techniques are a unique tool for: i) following the heat exchange process between the nanostructures and the substrate; ii) measuring small quantities of matter exploiting the shift of the pseudo-SAW frequency. In order to gain further understanding of the detailed mechanisms behind the experimental evidences and to develop devices exploiting optically driven thermo-mechanical transients, the solution of the full thermo-mechanical problem is required.

## 3. Ultrafast thermo-mechanics of nanostructured surfaces

After excitation by a single pump pulse, the thermo-mechanical evolution problem spans three time scales. In the first step, the laser short pulse heats the electron gas of the metallic nanodisks (subpicosecond time scale). In the second step, the hot electron gas thermalizes with the lattice (picosecond time scale). In the third step, two occurrences take place (a) a pseudo-surface acoustic wave (SAW) is launched in the system, finally transferring mechanical energy, δE$_{mech}$, to the Si bulk (b) the disks thermalize with the silicon substrate transferring heat, δQ, to the Si substrate (nanosecond time scale). This three-step sequence repeats itself upon arrival of a new laser pulse.

The absorption of a subpicosecond laser pulse induces an impulsive heating of the nanodisks occurring within the first few picoseconds. The physics is well modeled by the 2TM [Kaganov1957]:

$$C_e(T_e)\frac{\partial T_e}{\partial t} = P_p(t) - G \cdot (T_e - T_{Py}) + \vec{\nabla} \cdot (k_e \vec{\nabla} T_e) \qquad (4)$$

$$C_{Py}(T_{Py})\frac{\partial T_{Py}}{\partial t} = G \cdot (T_e - T_{Py}) + \vec{\nabla} \cdot (k_{Py} \vec{\nabla} T_{Py})$$

where T, k, and C indicate the temperature, thermal diffusion coefficient, and specific heat per unit volume, respectively, the reference to the electrons (e) or Permalloy (Py) being indicated by the subscript. $G$ is the electron-phonon coupling constant and $P_p(t)$ is the profile of the pulsed power per unit volume absorbed by the sample. Simulations of the time evolution of $T_e$ and $T_{Py}$ in the 30 nm-thick sample, show that the maximum temperature of the electron system, $T_e$=350 K, is reached 150 fs after laser excitation, whereas the thermalization with the lattice is completed on the picosecond time scale at a temperature of 313 K. The energy density absorbed by the sample, δU/V, is peaked within the nanodisks leaving the substrate temperature essentially unaltered, this selection being possible because of the difference in the optical penetration depths in Permalloy and Si. This occurrence gives rise to the onset of a mechanical and heat flux within the sample taking place on the nanosecond time scale.

As far as the thermal problem is concerned, the situation after ~5 ps is that of an isothermal nanodisk thermally linked to an isothermal Si substrate via an interface thermal resistivity $\rho_{th}=10^{-8}$ Km²W⁻¹ [Note1]. The thermal link translates in the following boundary conditions at the disk-substrate interface:

$$\widehat{\mathbf{n}}_{Py} \cdot k_{Py} \overrightarrow{\nabla} T_{Py} + (T_{Py} - T_{Si})/\rho_{th} = 0$$

$$-\widehat{\mathbf{n}}_{Si} \cdot k_{Si} \overrightarrow{\nabla} T_{Si} - (T_{Py} - T_{Si})/\rho_{th} = 0$$

(5)

$\widehat{\mathbf{n}}_{Py}$ and $\widehat{\mathbf{n}}_{Si}$ being the outward unit vector normal to the dot and Si boundary, respectively. The thermal resistivity is a macroscopic quantity accounting for phonons dispersion mismatch on the two sides of the interface and other eventual microscopic mechanisms limiting the heat flow between the metal disk and the Si substrate. The nanodisk temperature dynamics is readily accessible provided the Biot number Bi<<1, Bi being defined as Bi= h/(k$_{Py}$·ρ$_{th}$). Interpreting the ratio h/ρ$_{th}$ as an interface thermal conductivity, a value of Bi<<1 means that the disk remains isothermal during the thermal relaxation process to the substrate, implying a disk temperature evolution of the form ΔT(t)=ΔT(0)e$^{-t/\tau}$, where τ=hC$_{Py}$ρ$_{th}$ and ΔT(t)=T$_{Py}$(t)-T$_{Si}$(t). In the present case, assuming k$_{Py}$=20 WK⁻¹m⁻¹, we estimate Bi=0.15 for the 30 nm-thick sample and Bi=0.25 for the 50 nm-thick sample. Introducing the numbers for our samples we get τ=0.7 ns τ=1.1 ns, respectively, in qualitative agreement with the measurements reported in Figure 4. The time taken by the disk to thermalize with the substrate gives the time scale over which heat δQ is dissipated from the disk to the substrate. We stress that the thermal diffusion process within the nanostructure, responsible for double-exponential behaviors of the thermal relaxation [Hopkins2008], is confined within the first 5 ps and does not affect the reported measurements.

As far as the mechanics is concerned, the increase of the temperature of the periodic metallic nanodisks triggers a spatially modulated stress on the silicon surface. Such stress launches a pseudo-SAW of wavelength λ matching the nanodisks lattice periodicity. Oscillations arising in similar experiments have been tentatively interpreted in terms of eigenmodes of the single nanostructure or perturbation of free SAWs [Lin1993, Robillard2007, Robillard2008, Mante2008]. Only recently the full mechanical problem has been addressed beyond the simple perturbative approach, and pseudo-SAW solutions have been calculated [Nardi2009]. In a realistic physical scenario there is no distinction between the eigenmodes of nanostructures and SAWs, the solution of the elastic equation being a pseudo-surface acoustic wave partially localized on the nanostructures and radiating energy into the bulk. In Figure 5 we report the results of the calculations performed, considering the real dimensions of the samples. The large number of eigenmodes calculated, corresponding to coupling to different bulk modes, can be easily interpreted if the ratio between the energy content within 1 µm and within the entire cell, is reported as a function of the frequency. This ratio, named SAW-likeness coefficient account for the different localization on the surface of the calculated eigenmodes [Nardi2009]. Two Lorentzian curves emerge as the main contribution to pseudo-SAW solutions. In particular, the low energy one is related to sine-symmetry solutions, whereas the high energy one has cosine symmetry. The energy difference, of the order of 0.4 GHz, between the two modes is the phononic gap opening as a consequence of the periodicity of the system [Nardi2009]. The value of the pseudo-SAW eigenmodes, calculated within the latter theoretical frame, matches the experimental frequency ω, whereas the bulk mechanical energy content of the same eigenmode properly accounts for the energy radiated into the bulk, i.e. for the damping $\gamma$ reported in Figure 4.

## 4. Applications

The possibility to optically control the excitation of pseudo-SAW and thermal gradients in arrays of metallic nanostructures on substrates opens the way to fundamental applications in the field of hypersonic phononics and nanocalorimetry.

The gap in the phononic modes, reported in Figure 5, is the starting point to design and engineer surface waveguides and surface phononic cavities in the 10-100 GHz frequency range. In particular, the controlled removal of nanostructures adds localized modes within the phononic gap range. In Figure 6 we calculate the eigenmodes of long chain of 200 nm-wide, 50 nm-thick metallic nanostructures with periodicity of 1 µm, on top of a Si(100) surface. Upon removal of three metallic elements, a new mode, strongly localized within the cavity, appears at a frequency of ~4.45 GHz. This mode is prevented from propagating in the surrounding phononic crystal by the acoustic gap of the periodic structure. This simple element is the base of the development of waveguides and surface cavities, where the hypersonic acoustic pulses can be injected by excitation with ultrafast laser pulses. The accessibility of this kind of surface cavities constitutes a fundamental step toward the manipulation of molecules and small nanostructures by means of acoustic waves.

In addition, the sensitivity of the time-resolved techniques can be exploited to develop mass sensors with ps time-resolution, Considering that a difference of 10 ps in the oscillation period has been measured for the samples reported in Figure 4, we can easily estimate the sensitivity of these devices. The nanostructures volume difference in the two samples is $\Delta V=5 \cdot 10^{-17}$ cm$^3$, corresponding to a mass difference/disk of $\Delta m \sim 5 \cdot 10^{-16}$ g/disk. In the probe area the number of nanodisks is about 1250, giving an absolute mass variation of $\Delta m \sim 625$ fg. Moreover, sensitivity is dependent on several parameters such as the SAW wavelength, the unit cell filling factor and the disk's mass. Shorter SAW wavelengths imply higher surface confinement, hence higher surface sensitivity [Auld1973]. The proposed device periodicity can be scaled to tens of nanometers, thus enhancing the sensitivity.

Finally, the experimental scheme reported in this work proves very useful to perform specific heat measurements of mesoscale samples, that is for nanocalorimetry. When tackling the problem of measuring the specific heat of a small object a fast, non-contact probe is required. The speed requirement is dictated by the fact that the heat exchange between the sample and the thermal reservoir is proportional to the sample mass. For instance, in the present case, assuming knowledge of the interface thermal resistivity (this can be accessed with static thermoreflectance measurements), a sample mass of few femtograms leads to τ~1 ns, as discussed in the previous section. A contact probe of dimensions of the same order of magnitude of the dimensions of the sample affects the measurement in that one accesses the specific heat of the nano-sample and the probe itself. A non-contact probe would solve this problem. The scheme here outlined satisfies the above mentioned requirements: the Biot number is intrinsically small for metallic samples of thicknesses of tens of nm or less (see Section 3). This ensures a temperature decay time following a single exponential decay. The fast thermal flux can thus be probed with sub-ps time resolution by all-optical means. The ps time resolution allows probing, in principle, heat exchange processes in nanoparticles with dimensions of the order of few nanometers, where the relaxation time is expected to be in the sub-ns timescale, accordingly to the relation $\tau \propto h$, as discussed in Sec. 2.

## 5. Perspectives and conclusions

Until recently, the widespread application of devices based on time-resolved techniques has been hindered by two-main limitations: i) the weak variations of intensity of the probe pulse usually implies very long acquisition times; ii) the diffraction limit prevents from developing microscopy techniques with spatial resolutions smaller than hundreds of nanometers. Both these limitations can be overcome by the recent advances in ultrafast techniques.

Recent progresses in the control and synchronisation of the repetition rate difference of two femtosecond lasers can be exploited to make pump-probe measurements with a technique known as ASOPS (ASynchronous OPtical Sampling), without mechanical delay lines. This technique is based on two femtosecond lasers, stabilized at an offset repetition frequency Δf (typically in the range 1-10 kHz), much smaller than the repetition rate f (typically in the range 100–1000 MHz), This system produces two pulse trains in which successive pairs of pulses arrive at the sample with a delay that is linearly ramped from 0 to 1/f (which is the scanning window), in time steps whose amplitude depends on the frequency offset. Applications in which shot-noise limited relative reflection variations of $10^{-8}$ are attained by averaging pulse scans at a kHz rate, have been reported in the literature [Bartels2007]. The investigation of thermo-acoustic transients through reflectivity measurement, exploiting the ASOPS technique, would allow realizing an optical oscilloscope to follow in real time the strain waves propagation or fast thermal gradients induced by a short optical pump pulse.

An interesting approach to include lateral resolution in optical experiments is given by the Scanning Near-field Optical Microscopy (SNOM). This technique consists in sending light through an aperture much smaller than the light wavelength and then scanning the aperture relative to a sample at a distance (smaller than a wavelength) where the interaction of the near field protruding the aperture with the sample could be measured [Genet2007]. In this way, the interaction of the near field with the sample comes into play before the effects of diffraction, giving a sub-wavelength resolution. The integration of a SNOM with a femtosecond laser would allow adding time resolution to the lateral resolution provided by the SNOM. There are various possibilities, the simplest being to optically pump the sample externally and probe it through the near field emerging from the tapered fiber apex. In this way it is possible to follow the variation in the scattered near-field light interacting with the sample, collected inside the same fiber or externally, as a function of the delay between the near-field probe pulse and the far field pump pulse. With this technique it would be possible to follow the mechanical oscillation of a surface with a lateral resolution of the order of 50-100 nm. The near field scattering is provided, in this case, by the variation of the distance between the fiber apex and the surface due to the mechanical oscillations of the surface. The variation of the phase of the detected oscillation as a function of the spatial displacement provides a complete picture of the dynamics of the strain wave. In perspective, the combination of SNOM and ASOPS leads to the realization of optical oscilloscopes for the detection of ultrafast thermo-mechanical transients in matter with both temporal and spatial resolution.

In conclusion, we have reported on the experimental generation and detection of thermo-mechanical transients in matter, by means of femtosecond light pulses. Excitation of lattices of metallic nanostructures on surfaces allows investigating both the propagation of hypersonic pseudo-SAW in the substrate and the heat exchange process at the nanoscale. The phononic crystal behaviour of the system can be exploited to engineer surface cavities and acoustic waveguides in the 10-100 GHz frequency range, as the sizes are scaled down to tens of nanometers [Chao2005, Hurley2008, Siemens2009]. The possibility to combine the recently developed ASOPS and SNOM techniques allows designing real-time devices for mass and heat-exchange detection at the nanoscale with both sub-ps temporal and sub-µm spatial resolution

# References

[Arbouet2006] A. Arbouet, N. Del Fatti, and F. Vallée, "Optical control of the coherent acoustic vibration of metal nanoparticles", J. Chem. Phys. 124, 144701 (2006).

[Auld1990] B. Auld, *Acoustic Fields and Waves in Solids*, Vol.II (Krieger Publ., Malabar, FL, 1990).

[Bartels1999] A. Bartels, T. Dekorsy, and H. Kurz, "Coherent Zone-Folded Longitudinal Acoustic Phonons in Semiconductor Superlattices: Excitation and Detection", Phys. Rev. Lett. 82, 1044 (1999).

[Bartels2007] A. Bartels et al. "Ultrafast time-domain spectroscopy based on high-speed asynchronous optical sampling", Rev. Sci. Instrum. 78, 035107 (2007).

[Billaud2008] P. Billaud et al., "Correlation between the extinction spectrum of a single metal nanoparticle and its electron microscopy image", J. Phys. Chem. C 112, 978 (2008).

[Burgin2008a] J. Burgin, P. Langot, N. Del Fatti, F. Vallée, W. Huang, and M. A. El-Sayed, "Time-Resolved Investigation of the Acoustic Vibration of a Single Gold Nanoprism Pair", J. Phys. Chem. C 112, 11231 (2008).

[Burgin2008b] J. Burgin et al., "Acoustic Vibration Modes and Electron–Lattice Coupling in Self-Assembled Silver Nanocolumns", Nano Letters 8, 1296 (2008).

[Cecchini2005] M. Cecchini, V. Piazza, F. Beltram, D. G. Gevaux, M. B. Ward, A. J. Shields, H. E. Beere, and D. A. Ritchie, "Surface acoustic wave-induced electroluminescence intensity oscillation in planar light-emitting devices", Appl. Phys. Lett. 86, 241107 (2005).

[Cecchini2006] M. Cecchini, V. Piazza, G. D. Simoni, F. Beltram, H. E. Beere, and D. A. Ritchie, "Acoustic charge transport in a n-i-n three terminal device", Appl. Phys. Lett. 88, 212101 (2006).

[Chao2005] W. Chao et al., "Soft X-ray microscopy at a spatial resolution better than 15 nm", Nature 435, 1210 (2005).

[DelFatti1999] N. Del Fatti, C. Voisin, F. Chevy, F. Vallée, and C. Flytzanis, "Coherent acoustic mode oscillation and damping in silver nanoparticles", J. Chem. Phys. 110, 11484 (1999).

[deLima2005] M. M. de Lima, Jr., R. Hey, P. V. Santos, and A. Cantarero, "Phonon-Induced Optical Superlattice", Phys. Rev. Lett. 94, 126805 (2005).

[deLima2006] M. M. de Lima, Jr., M. van der Poel, P.V. Santos, and J. M. Hvam, "Phonon-Induced Polariton Superlattices", Phys. Rev. Lett. 99, 045501 (2006).

[Genet2007] C. Genet and T. W. Ebbesen, "Light in tiny holes", Nature 445, 39 (2007).

[Giannetti2007] C. Giannetti et al., "Thermomechanical behavior of surface acoustic waves in ordered arrays of nanodisks studied by near-infrared pump-probe diffraction experiments", Phys. Rev. B 76, 125413 (2007).

[Gundrum2005] B. C. Gundrum, D. G. Cahill, and R. S. Averback, "Thermal conductance of metal-metal interfaces", Phys. Rev. B 72, 245426 (2005).

[Hao2001] H.-Y. Hao and H. J. Maris, "Experiments with acoustic solitons in crystalline solids", Phys. Rev. B 64, 064302 (2001).

[Highland2007] M. Highland et al., "Ballistic-phonon heat conduction at the nanoscale as revealed by time-resolved x-ray diffraction and time-domain thermoreflectance", Phys. Rev. B 76, 075337 (2007).

[Hopkins2008] P. E. Hopkins, P. M. Norris, and R. J. Stevens, "Influence of inelastic scattering at metal-dielectric interfaces", J. Heat Trans. 130, 022401 (2008).

[Huang2004] W. Huang, W. Qian, and M. A. El-Sayed, "Coherent Vibrational Oscillation in Gold Prismatic Monolayer Periodic Nanoparticle Arrays", Nano Letters 4, 1741 (2004).

[Huang2005] W. Huang, W. Qian, and M. A. El-Sayed, "The Optically Detected Coherent Lattice Oscillations in Silver and Gold Monolayer Periodic Nanoprism Arrays: The Effect of Interparticle Coupling", J. Phys. Chem. B 109, 18881 (2005).

[Hurley2002] D. H. Hurley and K. L. Telschow, "Picosecond surface acoustic waves using a suboptical wavelength absorption grating", Phys Rev. B 66, 153301 (2002).

[Hurley2006] D. H. Hurley et al., "Time-resolved surface acoustic wave propagation across a single grain boundary", Phys. Rev. B 73, 125403 (2006).

[Hurley2008] D. H. Hurley, R. Lewis, O. B. Wright, and O. Matsuda, "Coherent control of gigahertz surface acoustic and bulk phonons using ultrafast optical pulses", Appl. Phys. Lett. 93, 113101 (2008).

[Huynh2006] A. Huynh et al., "Subterahertz Phonon Dynamics in Acoustic Nanocavities", Phys. Rev. Lett. 97, 115502 (2006).

[Kaganov1957] M. I. Kaganov, I. M. Lifshitz, and L. V. Tanatarov, Sov. Phys. JEPT 4,173 (1957).

[Landau1986] L. D. Landau and E. M. Lifshitz, *Theory of Elasticity* (Butterworth-Heinemann, Oxford, 1986).

[Lin1993] H.-Y. Lin et al., "Study of vibrational modes of gold nanostructures by picosecond ultrasonics", J. App. Phys. 73, 37 (1993).

[Lyeo2006] H.-K. Lyeo, and D.G. Cahill, "Thermal conductance of interfaces between highly dissimilar materials", Phys. Rev. B 73, 144301 (2006).

[Maris1998] H. Maris, *Picosecond Ultrasonics*, Scientific American, Vol. 278, p. 86, January 1998.

[Maznev2009] A.A. Maznev, "Laser-generated surface acoustic waves in a ring-shaped waveguide resonator", Ultrasonics 49, 1 (2009).

[Morozov2008] E. Morozov, Y. Laamiri, P. Ruello, D. Mounier, and V.E. Gusev, "Influence of the absorbed optical quanta energy on GHz ultrasound generation in GaAs", Eur. Phys. J. Special Topics 153, 239 (2008).

[Muskens2006] O. L. Muskens, N. Del Fatti, and F. Vallée, "Femtosecond Response of a Single Metal Nanoparticle", Nano Letters 6, 552 (2006).

[Nardi2009] D. Nardi et al. (2009, March). Pseudo-surface acoustic waves in hypersonic surface phononic crystals. arXiv cond-mat [Online]. 0904.0366v1. Available: http://arxiv.org/abs/0904.0366

[Nisoli1997] M. Nisoli et al., "Coherent acoustic oscillations in metallic nanoparticles generated with femtosecond optical pulses", Phys. Rev. B 55, R13 424 (1997).

[Note1] This value has been estimated as an average of the thermal resistivity values measured on similar interfaces (see E. Swartz and R. Pohl, "Thermal resistance at interfaces", Appl. Phys. Lett. 51, 2200 (1987), and R.J. Stoner and H. Maris, "Kapitza conductance and heat flow between solids at temperatures from 50 to 300 K", Phys. Rev. B 48, 16373 (1993), and P. E. Hopkins, P. M. Norris, and R. J. Stevens, "Influence of inelastic scattering at metal-dielectric interfaces", J. Heat Trans. 130, 022401 (2008)). At room temperature this value is always of the same order of magnitude irrespective of the hetero-junction materials.

[Perrin2006] B. Perrin, E. Péronne, L. Belliard, "Generation and detection of incoherent phonons in picosecond ultrasonics", Ultrasonics 44, e1277 (2006).

[Pezeril2008] T. Pezeril, F. Leon, D. Chateigner, S. Kooi, and K. A. Nelson, "Picosecond photoexcitation of acoustic waves in locally canted gold films", Appl. Phys. Lett. 92, 061908 (2008).

[Profunser2006] D. M. Profunser, O. B. Wright, and O. Matsuda, "Imaging Ripples on Phononic Crystals Reveals Acoustic Band Structure and Bloch Harmonics", Phys. Rev. Lett. 97, 055502 (2006).

[Robillard2007] J.-F. Robillard, A. Devos, and I. Roch-Jeune, "Time-resolved vibrations of two-dimensional hypersonic phononic crystals", Phys. Rev. B 76, 092301 (2007).

[Rogers2000] J. A. Rogers, A. A. Maznev, M. J. Banet, and K. A. Nelson, "Optical generation and characterization of acoustic waves in thin films", Annu. Rev. Mater. Sci. 30, 117 (2000).

[Rudolph2007] J. Rudolph, R. Hey, and P.V. Santos, "Long-Range Exciton Transport by Dynamic Strain Fields in a GaAs QuantumWell",Phys. Rev. Lett. 97, 047602 (2007).

[Scherbakov2007] A.V. Scherbakov et al., "Chirping of an Optical Transition by an Ultrafast Acoustic Soliton Train in a Semiconductor Quantum Well", Phys. Rev. Lett. 99, 057402 (2007).

[Siemens2009] M. E. Siemens, Q. Li, M. M. Murnane, H. C. Kapteyn, R. Yang. E. H. Anderson, and K. A. Nelson, "High-frequency surface acoustic wave propagation in nanostructures characterized by coherent extreme ultraviolet beams", Appl. Phys. Lett. 94, 093103 (2009).

[Stoner1993] R.J. Stoner and H. Maris, "Kapitza conductance and heat flow between solids at temperatures from 50 to 300 K", Phys. Rev. B 48, 16373 (1993).

[Sugawara2002] Y. Sugawara, O. B. Wright, O. Matsuda, M. Takigahira, Y. Tanaka, S. Tamura, and V. E. Gusev, "Watching Ripples on Crystals", Phys. Rev. Lett. 88, 185504 (2002).

[Tachizaki2006] T. Tachizaki, T. Muroya, O. Matsuda, Y. Sugawara, D. H. Hurley, and O. B. Wright, "Scanning ultrafast Sagnac interferometry for imaging two-dimensional surface wave propagation", Rev. Sci. Instr. 77, 043713 (2006)

[Taubert2007] R. Taubert et al., "Coherent acoustic oscillations of nanoscale Au triangles and pyramids: influence of size and substrate", New J. Phys. 9,  376 (2007).

[Thomsen1986] C. Thomsen, H. T. Grahn, H. J. Maris, and J. Tauc, "Surface generation and detection of phonons by picosecond light pulses", Phys. Rev. B 34, 4129 (1986).

[Vellekoop1998] M. J. Vellekoop, "Acoustic wave sensors and their technology", Ultrasonics 36, 1 (1998).

[Voisin2000] C. Voisin, N. Del Fatti,, D. Christofilos, and F. Vallée, "Time resolved investigation of the vibrational dynamics of metal nanoparticles", Appl. Surface Science 164, 131 (2000).

[vonGutfeld1966] R. J. von Gutfeld, and A. H. Nethercot, "Temperature dependence of heat-pulse propagation in sapphire", Phys. Rev. Lett. 17, 868 (1966).

# Figure Captions

**Figure 1**. Schematic of the experimental set-up used to generate and detect mechanical transients in nanostructured arrays, in the diffraction configuration. The pump beam is normal to the sample surface. The probe incident angle is smaller than 10°. The probe polarization is parallel to the plane of incidence.

**Figure 2**. A SEM image of the 50 nm-thick phononic crystal sample is reported.

**Figure 3**. Picture of the average heating process, related to the pump beam absorption modulated by the PEM.

**Figure 4**. Time-resolved measurements performed on (a) 30 nm-thick and (b) 50 nm-thick samples. The fit to the data (black line) of the sum of function (3) (red line) and a simple exponential decay ($e^{-t/\tau}$, blue dashed line), is reported. The red and blue curves have been scaled for graphical reasons. On the right, the top of the unit cell used in the finite element calculations is reported. The total depth of the cell used in the calculations is 100 μm. The dot thermal expansion profile at t=5 ps (top) and the displacement profile of the impulsively excited pseudo-surface acoustic wave (bottom), are shown.

**Figure 5**. Acoustic eigenmodes calculated considering nanodots of diameter 400 nm and height 50 nm. The SAW-likeness coefficient of the calculated eigenmodes is reported as a function of the frequency, outlining two Lorentzian curves corresponding to sine and cosine symmetry solutions. The 0.4 GHz phononic gap between the two modes is reported. The pictures represent the total displacement (normalized color scale) and the vectorial displacement (arrows) of the pseudo-SAW solutions, evidencing radiation of mechanical energy into the substrate.

**Figure 6**. Calculated pseudo-SAW eigenmodes of long chain (19 elements) of 200 nm-wide, 50 nm-thick metallic nanostructures with periodicity of 1 μm, on top of a Si(100) surface. Upon removal of three metallic elements, a new mode strongly localized within the cavity appears at a frequency of 4.45 GHz.

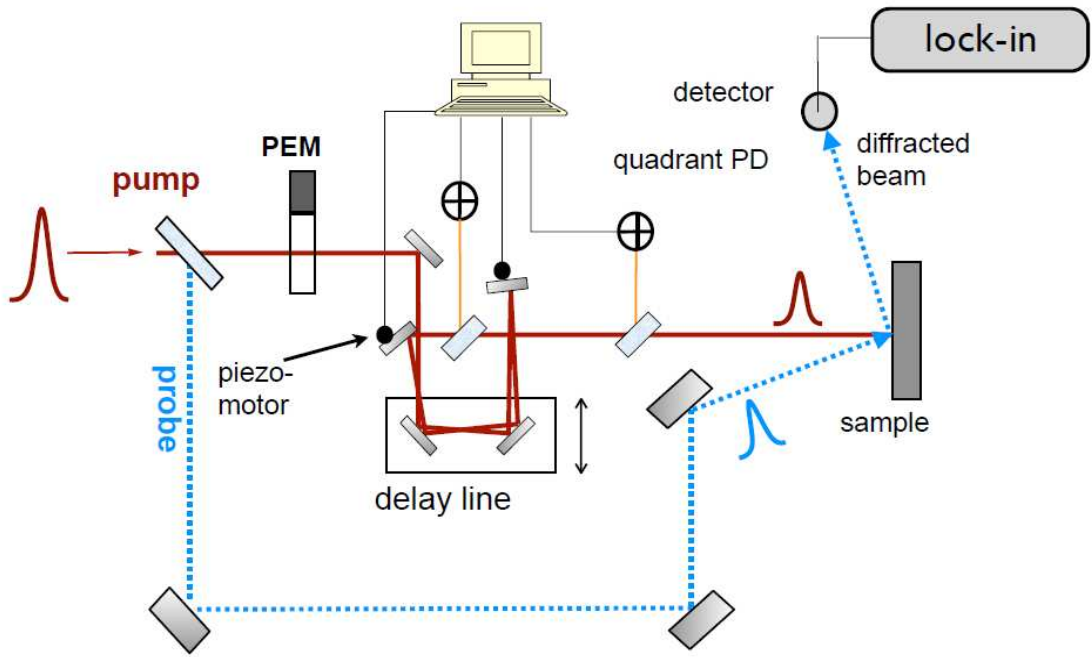

Figure 1

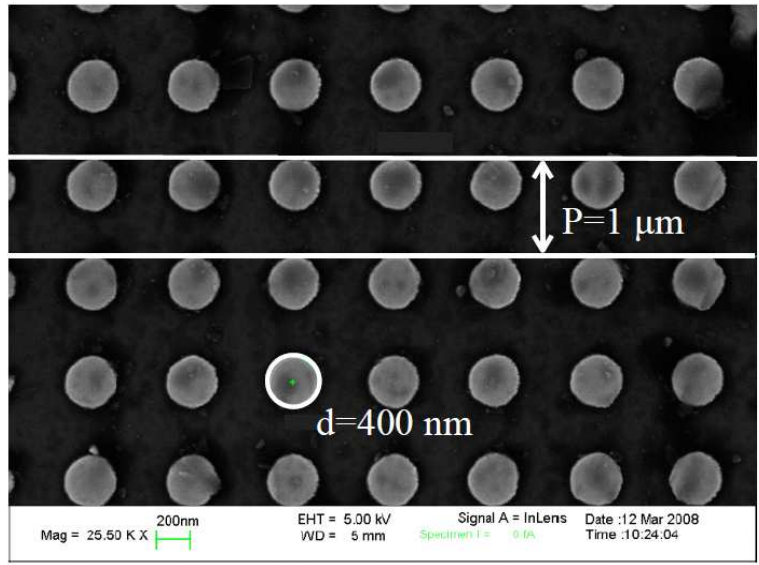

Figure 2

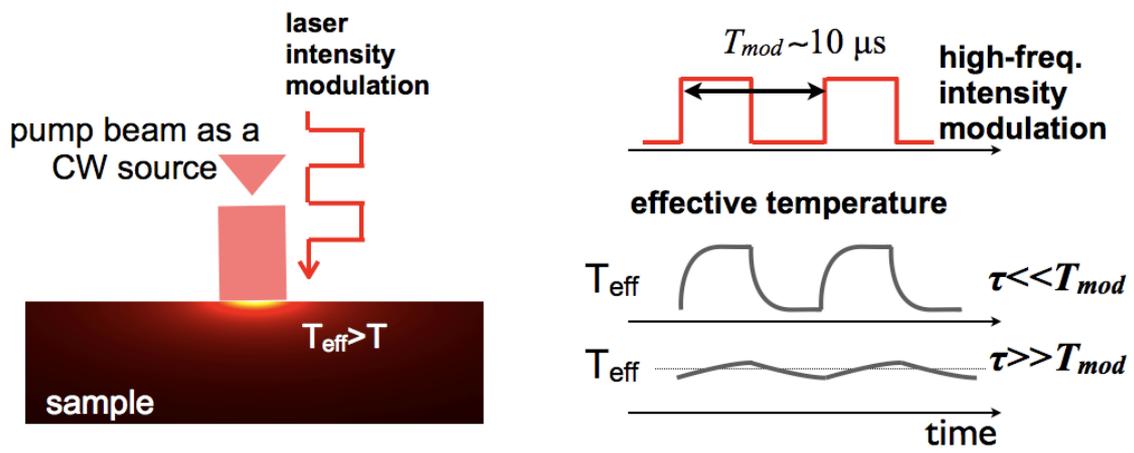

Figure 3

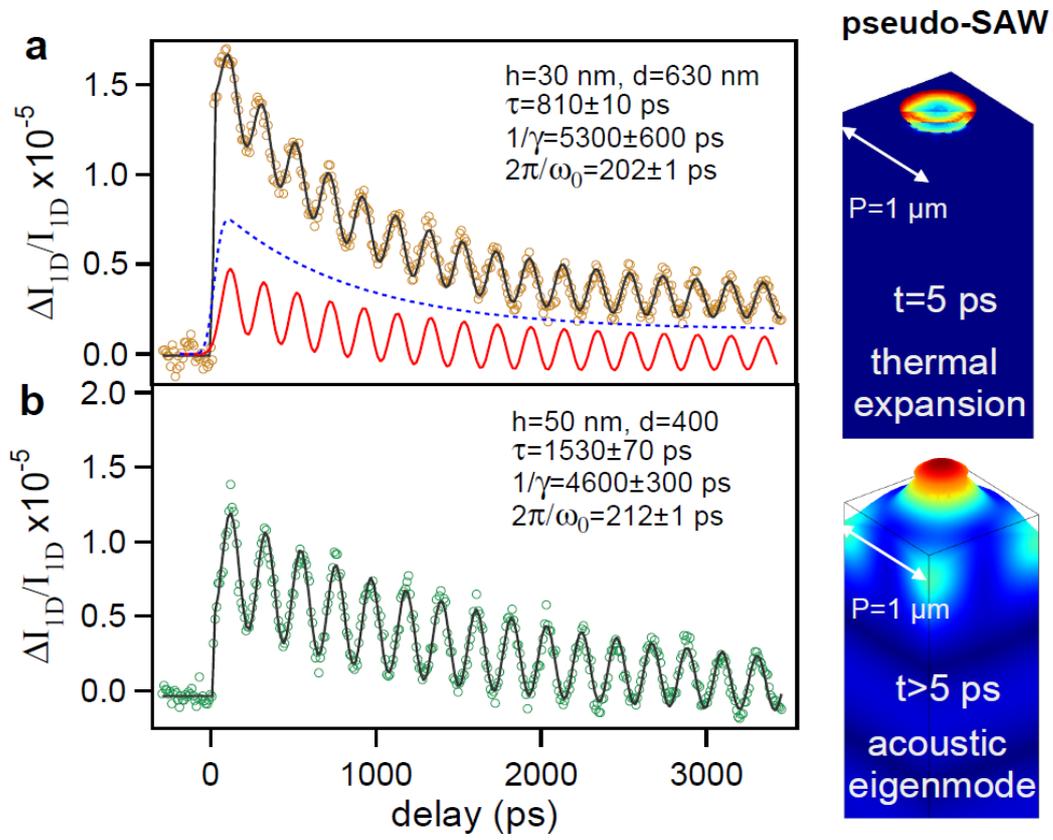

Figure 4

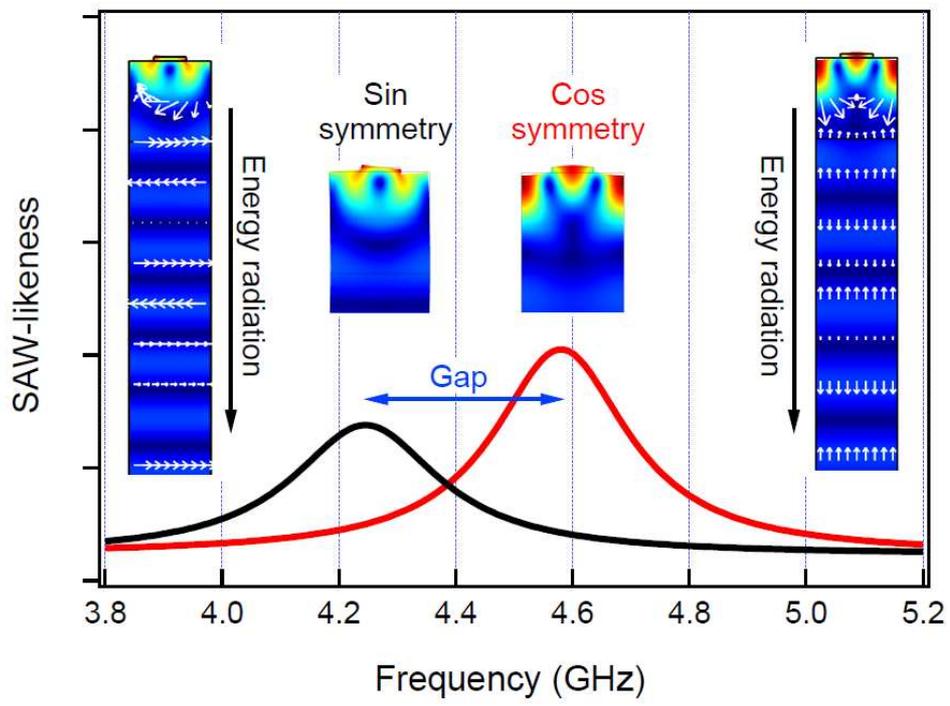

Figure 5

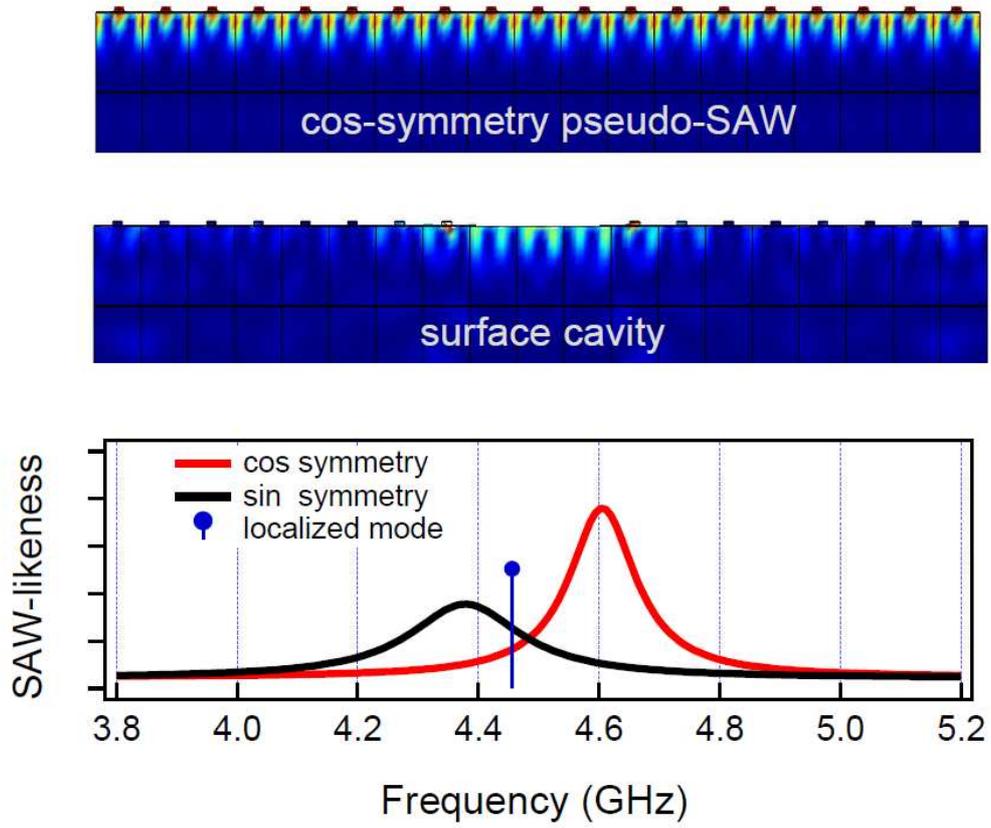

Figure 6